\documentclass [twocolumn]{revtex4-1}
\usepackage{amssymb,amsfonts,amsmath}
\usepackage{graphicx}
\usepackage{multirow}
\usepackage{longtable}
\usepackage{setspace}
\usepackage{epsfig,latexsym,graphicx,bm,pifont}
\usepackage{color}
\providecommand\bcdot{\boldsymbol{\cdot}}
\providecommand\bnab{\mbox{\boldmath $\nabla$}}

\providecommand\bcdot{\boldsymbol{\cdot}}
\providecommand\bnab{\mbox{\boldmath $\nabla$}}

\providecommand\btau{\mbox{\boldmath $\tau$}}

\begin{document}

\title{Dynamical Self-regulation in Self-propelled Particle Flows}

\author{Arvind Gopinath$^{1}$, Michael F. Hagan$^{1}$, M. Cristina
Marchetti$^{2}$ and Aparna Baskaran$^{1}$}

\affiliation{$^{1}$Martin Fisher School of Physics, Brandeis University, Waltham, MA, USA.\\
$^{2}$Physics Department and Syracuse Biomaterials Institute, Syracuse University, Syracuse, NY 13244, USA.\\}

\begin{abstract}
We study a continuum model of overdamped self-propelled particles with an aligning interaction in two dimensions.
By combining analytical and numerical work ,we map out the phase diagram for generic parameters.
We find that the system self-organizes into two robust
structures in different regions of parameter space: solitary waves of ordered swarms moving through a low
density disordered background, and stationary asters. The self-regulating nature of the flow yields phase separation, ubiquitous in this class of systems, and controls the formation of solitary waves. Self-propulsion and the associated active convection play a crucial role in aster formation. A new result of our work is a phase diagram that displays these different regimes in a unified manner.
\end{abstract}

\date{\today}
\maketitle
\section{Introduction}

Active fluids are composed of interacting self-propelled particles that individually consume energy and collectively generate large scale motion. Examples span many
length scales, ranging from animal herds \cite{parrish},
schools of fish \cite{fish}, bird flocks \cite{starling} and insect swarms
\cite{insects}, to bacterial colonies \cite{B_Subtilis, Myxobacteria,
Wu_Libchaber} and the cell cytoskeleton \cite{MackintoshScience07}. In this paper, we consider the overdamped dynamics of a collection of self-propelled particles subject to local aligning interactions. The model is relevant to various experimental systems, including motility assays \cite{Schaller}, suspensions of cytoskeletal filaments~\cite{nedelec}, self-chemotactic bacteria such as E-coli in convection-free geometries~\cite{SilberzanPNAS}, and inanimate systems such as vibrated granular monolayers \cite{grains} and chemically driven  nano-rod suspensions \cite{SPR}.

The goal of the study is to identify universal hydrodynamic mechanisms for the emergence of complex structures in systems exhibiting collective motility. To this end, we focus on a simple and generic macroscopic description in terms of a conserved density field and a collective velocity or polarization field. The continuum model is expected to capture the behavior of the system on length scales large compared to the size of the individual units and time scales long compared to the microscopic interaction times and the frictional time scale set by the medium. Such a description was first considered in the pioneering work of Toner and Tu~\cite{TonerTu95,TuToner98,TTRreview05} who found that self-convection inherent to active particle flows  stabilizes long range order in two dimensions. In this work we study this generic coarse-grained description analytically and numerically, map out emergent inhomogeneous structures, and identify the mechanisms underlying their formation.

\begin{figure}
\includegraphics [width=8cm] {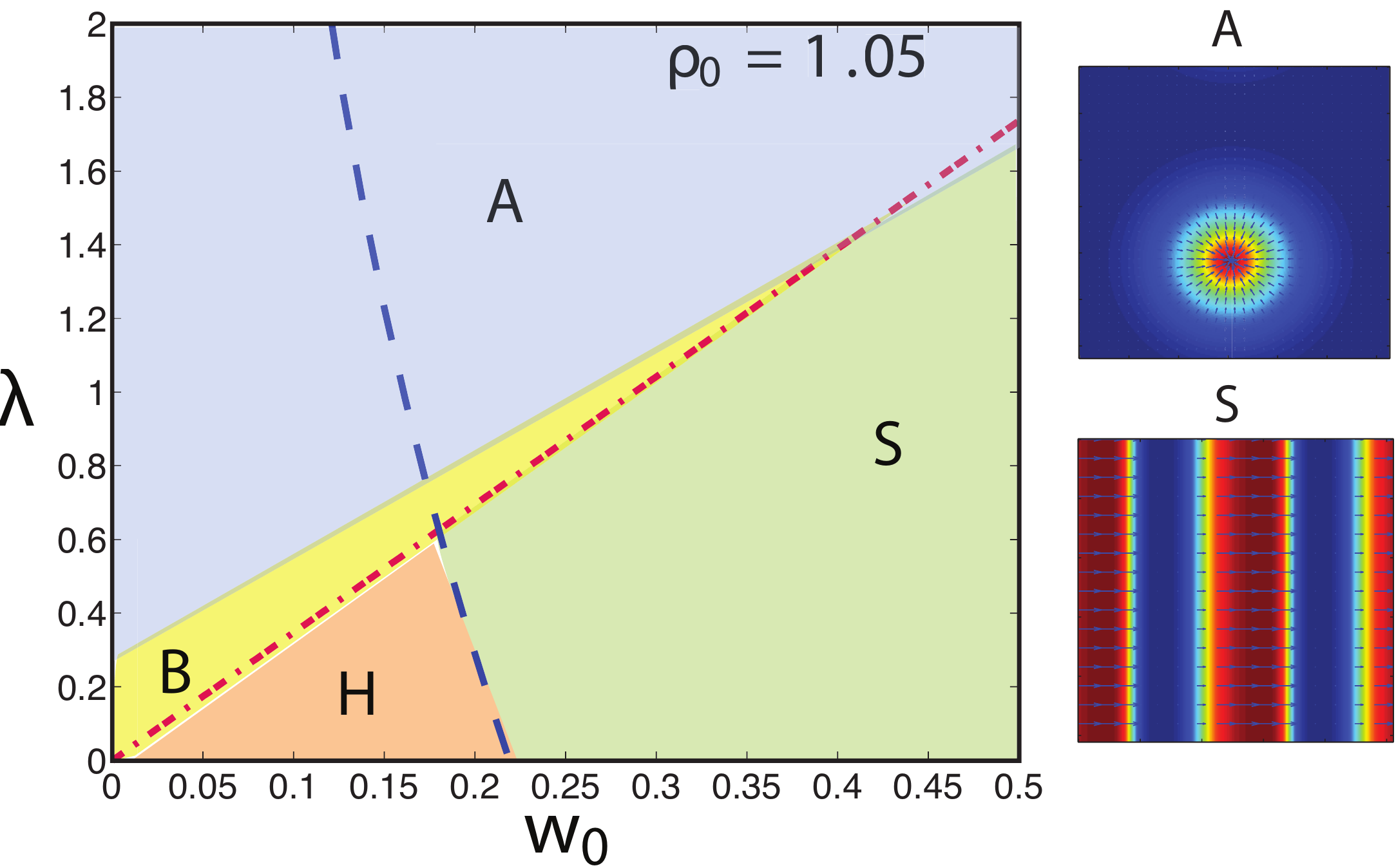}
\caption{ (Color online) {\it Left Panel} : Phase diagram in the $(\lambda,w_0)$ plane for $\rho_0=1.05$. The dashed (blue online) and dashed-dotted (red online) lines are the neutral stability curves for the L (Eq.~\eqref{L-inst}) and T (Eq.~\eqref{T-inst}) modes, respectively. The shaded regions describe the  stable states obtained numerically: a homogenous polar state (H), a regime of  propagating solitary stripes (S), a regime of   stationary asters (A),  and moving localized polar clusters (B). The domain size is (in dimensionless units)  $128 \times 128$ and the equations were integrated up to $5 \times 10^{4}$ diffusion times. The moving polar clusters are a result of finite system size and the finite time of integration. {\it Right Panel} : Snapshot of a propagating stripe (bottom) and a stationary aster (top), the color scheme denotes density (increasing from blue to red).}
\label{phase-diagram}
\end{figure}

We study the system as a function of two parameters, the self-propulsion speed $w_0$ of the active particles and a parameter $\lambda$ that incorporates the effect of inter-particle interactions. The theoretical model of Toner and Tu~\cite{TonerTu95,TuToner98,TTRreview05} yields a mean field transition from a disordered state to an ordered ``polar" state (i.e., a state with nonzero mean velocity) that is controlled by the density of the active units. This density is not, however, an external control parameter as in conventional equilibrium phase transitions, but rather is convected by the order parameter. This coupling renders the dynamics of the system self regulating, in that the state of the system is determined by the interplay between particle convection and tendency to local alignment, rather than by an externally controlled density of  active particles.

Our main result is the phase diagram shown in Fig.~\ref{phase-diagram}, that displays the  various dynamical states obtained by varying $w_0$ and $\lambda$, for a fixed density above the mean-field order-disorder transition. As expected, the homogeneous polar state (H) is unstable in most of the parameter space. It is replaced by one of two robust structures in different regions of parameters. The first is a state of propagating solitary waves consisting of high density ordered swarms moving through a low density disordered background (see Fig.~\ref{phase-diagram}, Region (S)). The second is  a stationary aster (see Fig.~\ref{phase-diagram}, Region (A)).

Density waves of the type found here are ubiquitous in bacterial systems~\cite{SilberzanPNAS,Adler,Park,Salman,Austin,Magnetotactic,BsubtilisStripes} and have also been observed in dense motility assays of short actin filaments~\cite{Schaller}. Theoretical studies showing the emergence of traveling wave structures have included diffusion models with chemotactic gradients~\cite{KellerOld,BrennerOld,ArmitageRev,SilberzanPLoS}, numerical simulations of agent based models, such as the Vicsek model~\cite{Chate} and coarse grained theories~\cite{bertin2006,bertin2009,Mishra2010}. In particular, Bertin et al~\cite{bertin2009} have recently pointed out that the traveling density stripes are solitary waves, rather than a nonequilibrium pattern of the system. In this work, we show that these density waves are an inevitable consequence of the self-regulating nature of self-propelled particle flows and can be viewed as a coexistence between two stable phases of the system, namely a high density ordered swarming state and a low density disordered state.

Asters are ubiquitous in cell biology in processes such as the formation of the mitotic spindle \cite{spindlelit,spindlelit2}. They also occur in {\em in-vitro} suspensions of cytoskeletal filaments and motor proteins \cite{nedelec,nedelec2,surrey}. Theoretical models of mixtures of cytoskeletal filament and  motor proteins do indeed yield aster formation ~\cite{Kardar,Sankararaman,Joanny,Aranson,Zimmermann,MarchettiTBL,surreyTheory}, arising from the dynamics of the motor proteins and/or the flow of the solvent. Asters have not, however, been obtained before in Toner-Tu continuum models of self-propelled particles because they only occur for larger value of the effective nonlinearities than considered in previous work. In addition, our work identifies a universal hydrodynamic mechanism for  aster formation in the change in sign of the effective nonequilibrium compressibility of the system, combined with the active self convection. 

The layout of the paper is as follows. First, we review the hydrodynamic theory  and describe the key features that control the emergent structures. Then, we carry out a linear stability analysis of the homogeneous swarming state. Next, we report the results of a numerical solution of the nonlinear deterministic equations and discuss the mechanisms underlying the formation of the propagating density waves and stationary asters. Finally, we conclude with a discussion aimed at placing this work in the context of the existing vast literature on active polar fluids. 

\section{The Continuum Model}

We model the overdamped dynamics of a collection of self-propelled particles. The only conserved field is the number density $\rho({\bf r},t)$ of active particles. In addition, to describe the possibility of states with polar orientational order or collective motility, we consider the dynamics of a vector field, $\bm\tau({\bf r},t)=\rho({\bf r},t){\bf P}({\bf r},t)$, that represents a polarization density.  Here, $\mathbf{P}(\mathbf{r},t) $ is
an order parameter for polar orientational order. Its magnitude measures the amount of orientational order and its
direction represents the Goldstone mode associated with the spontaneously broken rotational
symmetry in the swarming state. The dynamical equations associated with these quantities were first constructed phenomenologically by Toner and Tu \cite{TonerTu95} and have more recently been derived from various microscopic models~\cite{Aranson, MarchettiTBL, Ahmadi, Aparna, bertin2006,bertin2009}. They are given by
\begin{equation}
\label{rho-eq}\partial_t\rho=-\bm\nabla\cdot(w_0\bm\tau-D\bm\nabla\rho)\;,
\end{equation}
\begin{eqnarray}
\label{tau-eq-2}\partial_t\bm\tau+&&\lambda_1(\bm\tau\cdot\bm\nabla)\bm\tau=-\left[a_2(\rho)+a_4|\bm\tau|^{2}\right]\bm\tau+K\nabla^2\bm\tau\nonumber\\
&&-w_1\bm\nabla\rho+\frac{\lambda_3}{2}\bm\nabla |{\btau}|^2+\lambda_2{\btau}(\bm\nabla\cdot{\btau})\;.
\end{eqnarray}
The density equation, Eq.~\ref{rho-eq},  is a conservation law with a mass flux controlled by the sum of
convection  of the active particles at the self-propulsion speed $w_0$ and diffusion at rate $D$.
The structure of the polarization equation, Eq.~\ref{tau-eq-2}, reflects the fact that $\bm\tau$ plays a dual role: on one hand  $\bm\tau/\rho$ is the orientational order parameter of the system, on the other $w_0\bm\tau/\rho$ represents the mean flow velocity.
The first two terms on the right hand side of Eq.~\ref{tau-eq-2} control the mean-field continuous order-disorder transition to a state of collective motility, with $a_2(\rho)$ a parameter that changes sign at a characteristic density $\rho_c$, and $a_4(\rho)>0$.  The term proportional to $\lambda_1$ describes self-convection. It is the analog of the finite Reynolds number convective nonlinearity in the Navier-Stokes equation. Unlike the fluid, this overdamped system does not possess Galileian invariance as the particles are moving relative to a medium. As a result, $\lambda_1\not=1/\rho$. The lack of Galileian invariance also allows other convective terms proportional to $\lambda_2$ and $\lambda_3$ appearing on the right hand side of the equation. In contrast to the term proportional to $\lambda_1$, these terms also have an equilibrium interpretation and can be present in an equilibrium polar or ferroelectric fluid \cite{MarchettiKung}. The term proportional to $w_1$ is essentially a pressure gradient and is unique to systems with polar symmetry. Before proceeding further we point out the two crucial properties of the above equations that control the formation of emergent structures in this system.

\textbf{Dynamical Self-regulation.}  It is useful to compare  Eq.~\ref{tau-eq-2} to that for the order parameter of an equilibrium lyotropic polar liquid crystal such as a smectic C Langmuir monolayer \cite{degenneBook}.  The important difference is that in the equilibrium system the density controlling the order-disorder transition is externally tuned. In the case of a collection of self-propelled particles, the density is \emph{convected by the order parameter} itself through the $w_0$ term in Eq.~\ref{rho-eq}. In this sense the amount of order is itself regulated by the dynamics of the system and this coupling is rendered non-local by the convective terms. This is a crucial feature of the dynamics of self-propelled systems and plays a central role in controlling the formation of emergent structures.

\textbf{Negative effective compressibility.} Two terms on the right hand side of Eq.~\ref{tau-eq-2} that are along the direction of the spatial gradient, namely $w_1\bm\nabla\rho-\frac{\lambda_3}{2}\bm\nabla|{\btau}|^2$ represent  pressure gradients, with $w_1\rho-\frac{\lambda_3}{2}|{\btau}|^2$ the effective pressure. The first term is  the ideal gas part of the pressure. The second term arises due to the intrinsic tendency of polar systems to splay \cite{MarchettiKung}. For $\lambda_3>0$, as obtained from all microscopic derivations~\cite{Mishra2010,bertin2009}, it describes a lowering of the pressure due to ordering of the velocities. When $\lambda_3$ becomes large the system develops an effective negative compressibility. This is the central property that controls the physics of this system in the interaction dominated regime.

Given the large number of parameters in the hydrodynamic equations, Eq.~\ref{rho-eq}-~\ref{tau-eq-2}, we simplify them as follows.
We choose
 $a_{2}=\nu\left(1-\rho/\rho
_{c}\right)$ and $a_{4}=\left(\nu/\rho^2\right)\left(1+\rho /\rho _{c}\right)$, with $\nu$ a characteristic kinetic frequency. This yields a continuous
(in dynamical systems terms - a supercritical pitchfork) mean-field
phase transition from an isotropic ($\btau=\mathbf{0}$) to a homogeneous, polar or
swarming state ($|\btau| > 0)$ at the critical density $\rho =\rho _{c}$ and ensures
that $|\tau|/\rho_0\rightarrow 1$ for $\rho \gg  \rho_c$.
We further assume $D=K$. In the following we measure time in units of $\nu^{-1}$ and  lengths
in units of $ (D/\nu)^{1 \over 2}$.
Without loss of generality, we also set the critical density $\rho _{c}=1$.
To reduce the number of independent parameters the numerical work is carried out for   $w_1=w_0$ and
$\lambda_1=\lambda_2=\lambda_3=\lambda$.
The equations then involve three parameters: (1) the
mean density of the system, $\rho _{0}$, which determines
the distance from
the order-disorder transition, (2) the convective velocity $w_{0}
$, and (3) $\lambda $ which
is controlled by the strength of interparticle interactions.
The
hydrodynamic equations in dimensionless simplified
variables are given by
\begin{subequations}
\begin{gather}
\partial _{t}\rho =-\bnab\bcdot(w_{0}\btau-\bnab\rho )\;,  \label{rho-eq-s}\\
\partial _{t}\btau=-(a_{2}+a_4\btau ^{2})\btau-w_{0}\bnab
\rho +\nabla ^{2}\btau\nonumber\\
+\lambda\left( \tau _{\alpha }\bnab\tau _{\alpha }+\btau\bnab\cdot \btau-\btau
\bcdot\bnab\btau\right)\;,  \label{tau-eq-s}
\end{gather}
\end{subequations}
where $w_0$ and $\lambda$ are now dimensionless.

\section{Linear stability analysis}

In this section, we examine the linear stability of the homogeneous solutions to Eqs.~\eqref{rho-eq} and \eqref{tau-eq-2}. We use dimensionless variables, but to highlight  the role of each term in hydrodynamic equations, we retain the different parameters $w_i$ and $\lambda_i$.
The continuum equations have two homogeneous, stationary solutions, both with $\rho=\rho_0={\rm constant}$: an isotropic state with  $\btau=0$ for $\rho_0 <\rho _{c} \equiv 1$ and a polar state  with $\btau\not=0$ for
$\rho_0 >\rho _{c} $. We focus here on the polar (or collective motility) state.
Without loss of generality, we
take the direction of polarization to be the $x$ axis of our coordinate system.
The homogeneous polar state is then characterized by
$\btau=\tau _{0}\mathbf{\hat{x}}$ with $\tau _{0}=\rho _{0}\sqrt{\left(
\rho _{0} - \rho _{c}\right) / \left(\rho _{0} + \rho _{c}\right)}$.

We now examine the stability of this state
 to  small amplitude perturbations by letting
$\rho =\rho _{0}+\delta \rho \left( \mathbf{r},t\right) $, $\btau=
\btau_{0}+\delta \tau \left( \mathbf{r},t\right) \mathbf{\hat{x}}+\delta \tau
_{\bot }\left( \mathbf{r},t\right)\mathbf{\hat{y}}$. Introducing the Fourier representation $\widetilde{x}
\left( \mathbf{q},t\right) =\int d\mathbf{r}e^{i\mathbf{q}\cdot \mathbf{r}
}x\left( \mathbf{r},t\right)$, the linearized equations are given by
\begin{subequations}
\begin{gather}
\partial _{t}\delta \widetilde{\rho }=-q^{2}\:\delta \widetilde{\rho }
+iw_{1}(q_\Vert\:\delta \widetilde{\tau }+q_\bot\:\delta
\widetilde{\tau }_{\perp })\;,  \label{lin-rho}\\
\partial _{t}\delta \widetilde{\tau }=(\alpha _{1}+iw_{1}q_\Vert
)\:\delta \widetilde{\rho }-(\alpha _{2}+i\overline{\lambda} \tau _{0}q_\Vert+q^{2})\:\delta \widetilde{\tau }\nonumber\\
-i\lambda_2 \tau _{0}q_\bot\:\delta \widetilde{\tau }_{\perp }\;,
\label{lin-tau}\\
\partial _{t}\delta \widetilde{\tau }_{\perp }=iw_{0}q_\bot\:\delta
\widetilde{\rho }-
i\lambda_3 \tau _{0}q_\bot\: \delta \widetilde{\tau}
+(i\lambda_1
\tau _{0}q_\Vert-q^{2})\: \delta \widetilde{\tau }_{\perp } \;, \label{lin-tau-perp}
\end{gather}
\end{subequations}
with $\overline{\lambda}=\lambda_3+\lambda_2-\lambda_1$, $q_\Vert=q\cos\theta$, $q_\bot=q\sin\theta$ and $\theta $
the angle between the wavevector $\mathbf{q}$ and the direction of
broken symmetry, $\mathbf{\hat{x}}$.
Also, we have defined  $\alpha _{1}\equiv -\tau _{0}\left( {\frac{{\partial
a_{2}}}{{\partial \rho }}}-{\frac{a_{2}^{0}}{a_{4}^{0}}}{\frac{{\partial
a_{4}}}{{\partial \rho }}}\right) $, and $
\alpha _{2}\equiv -2a_{2}^{0}$, with  $\alpha _{1}\geq 0$ and $\alpha _{2}\geq 0$ for $\rho
_{0}>\rho _{c}$. 
We  seek
solutions of the form
$\delta \widetilde{\rho },\delta \widetilde{\btau }\sim e^{s_{\alpha }\left( \mathbf{q}\right) t}$.
The homogeneous state will then be stable provided  ${\mathrm{Re}}[s_\alpha(\mathbf{q})]<0$ for all eigenvalues.
The linear stability of the homogeneous polar state has been discussed elsewhere \cite{Mishra2010,bertin2009} and will be only summarized here with focus on the aspects relevant for emergent structures.
The physics is highlighted by examining the special cases of wavevector ${\bf q}$
 along the direction of broken symmetry ($\theta =0$) and normal to it ($\theta=\pi/2$).

\textbf{Convection mediated density instability.}   When $\theta =0$ and $q=q_\Vert$, the  fluctuations $\delta \tau _{\bot }$ decouple and are always stable.
The coupled
equations for the density and magnitude fluctuations give the dispersion relations of the other two  modes.
In the long wavelength limit $q \rightarrow 0$ one finds that these modes are unstable for
\begin{equation}
 1+{\frac{w_{0}w_1}{\alpha _{2}}}-w_{0}^{2}{
\frac{\alpha _{1}^{2}}{\alpha _{2}^{3}}}-\overline{\lambda} \tau _{0}w_{0}{\frac{\alpha
_{1}}{\alpha _{2}^{2}}} <0\;.\label{L-inst}
\end{equation}
This instability is intimately related to the self regulating nature of the dynamics. It arises from the density dependence of the local  tendency of the system to build up polar order, $a_2(\rho)$, and the fact that this density dependence is rendered nonlocal by the convective coupling of density to $\btau$ proportional to $w_0$ and $w_1$. The location of the longitudinal  instability line defined by Eq.~\eqref{L-inst} in the $(\lambda,w_0)$ plane is shown by the dashed (blue online) line in Fig.~\ref{phase-diagram} for a fixed value of density. Alternatively, one can identify a density $\rho _{R}(\lambda ,w_{0})$ such that  magnitude fluctuations in the order parameter destabilize the
homogeneous polar state for $\rho_R>\rho_c$, leading to the onset and growth of density inhomogeneities.This longitudinal instability has been discussed previously in Refs.~\cite{bertin2009} and \cite{Mishra2010} and will be referred to here as the convection mediated density instability. We want to emphasize two features of this instability relevant for this work : i) this instability is mainly controlled by the third term in Eq.~\eqref{L-inst} and as such is model independent and arises purely due to the convective coupling between the collective velocity and density; ii) as the order-disorder transition is approached from above
$\alpha_2\rightarrow 0$ and $w_{0c}\sim(\rho_0-\rho_c)^{1/2}$, i.e., near $\rho_c$ the mean-field ordered state becomes unstable for vanishingly small $w_0$.

\textbf{Splay induced negative compressibility.}  We now consider the dynamics of fluctuations in the direction orthogonal to the
polarization, i.e., for $\theta =\frac{\pi }{2}$ or $q=q_\bot$. In this case  the modes are all stable and diffusive near the isotropic-polar mean-field transition~\cite{Mishra2010}. Far from the transition fluctuations in $\delta\tau$ decay rapidly and can be
eliminated in favor  of  $\delta \rho $ and $\delta
\tau _{\bot }$ by letting
$\delta \tau \approx \left( {\frac{{\alpha _{1}}}{{\alpha _{2}}}}\right)
\delta {\rho }-\left( {\frac{{iq\lambda_3 \tau _{0}}}{{\alpha _{2}}}}\right)
\delta \tau _{\perp }$. Substituting this in
Eqs. ~\eqref{lin-rho} and \eqref{lin-tau-perp}  we find that $\delta\tau_\bot$  fluctuations, which in this case describe splay deformation of the order parameter, are unstable for
\begin{equation}
w_{1}-{\lambda_3 \tau _{0}{\frac{\alpha _{1}}{\alpha _{2}}}} <0\;.
\label{T-inst}
\end{equation}
This splay instability
is controlled by nonlinear couplings proportional to   $\lambda_3$ and occurs when the effective compressibility of the system as determined by the third and fourth terms on the right hand side of Eq.~\eqref{tau-eq-2} changes sign.The parameter  $\lambda_3$ is in turn determined by interactions in the system~\cite{bertin2009,Mishra2010}, hence the precise location of the instability line  is model dependent.

To summarize,
we have identified two  mechanisms that render the system linearly unstable and lead to growth of
density fluctuations.
In the next section, we go beyond the linear stability analysis and use detailed numerical computations to characterize the spatially (and possibly temporally) inhomogeneous states
that replace the homogeneous solution in the unstable region of parameters.

\section{Emergent Structures}

To study the emergent structures arising from the nonlinear model, Eqs.~\eqref{rho-eq-s} and \eqref{tau-eq-s} were solved numerically
in two spatial dimensions using an explicit (FTCS) scheme and a semi-implicit Fourier-Galerkin scheme.
In most calculations the system was started at $t=0$ in  a homogenous, non-polar (disordered) state, with small amplitude, uniformly distributed, zero-mean noise. Local initial density perturbations were chosen to be less than $3$\% of the mean density.

The results are summarized in the phase diagram shown in Fig~\ref{phase-diagram} and discussed briefly in the introduction.  For large values of activity the homogenous polar state (H) is unstable and  two steady, inhomogeneous states are obtained: (i) propagating stripes comprised of ordered swarms moving through a disordered background, when active convection exceeds the strength of nonlinearities  $w_0 / \lambda \gg 1$, and (ii)  a stationary aster
when non-linear effects  dominate convection $\lambda / w_0 \gg 1$. In this section, we discuss  the mechanisms underlying the formation of these two emergent structures.

\begin{figure}
\includegraphics[width=7cm]{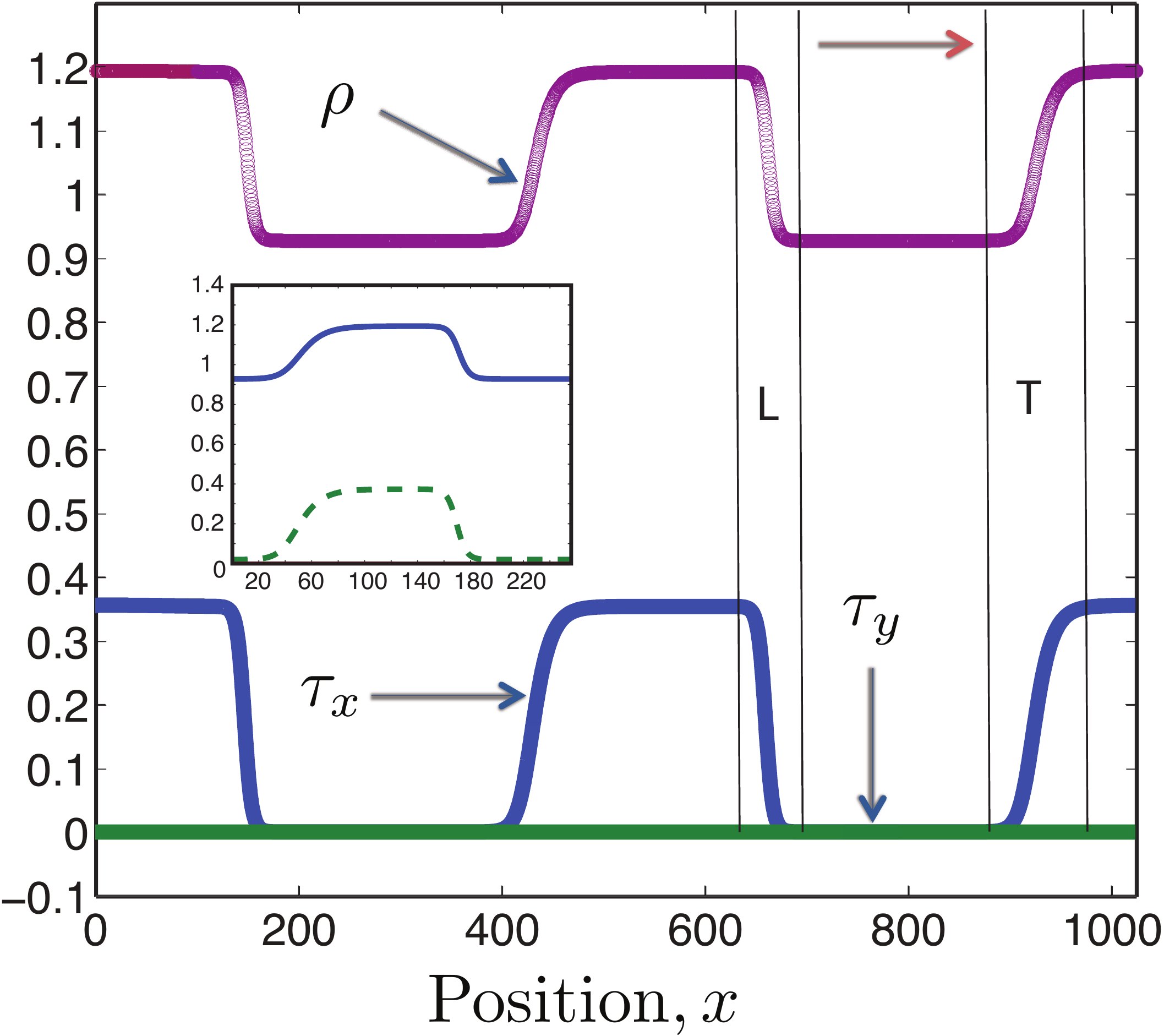}
\caption{Profiles of $\rho$, the
density (pink in online version) and $\tau_{x}$ and $\tau_{y}$ the polarization fields (blue and green online) are plotted
for a striped state when $w_0=0.4$, $\lambda=0$ and $\rho_0=1.05$. The system size is $1024 \times 32$ and results shown are obtained after integrating to $10^{4}$ diffusion times. The arrow on the top right (red online) denotes the direction of motion of the stripes The vertical lines demarcate the trailing (T) and leading (L) edges of the stripe. The inset shows the details of the profile of density (solid line, blue online)  and polarization (dashed line, green online) for one stripe.}
\label{stripe-structure}
\end{figure}

\subsection{Solitary waves}

We begin by considering the dynamics of the system in the convection dominated regime. For values of $w_0$ above the critical value for the onset of the convective density instability, defined by Eq.~\eqref{L-inst}, an initially homogeneous state develops hot-spots of high density that are then  convected throughout the system due to the coupling between
$\btau$ and $\rho$. These high density regions reorganize and  grow in the direction lateral to their motion
due to diffusion, resulting in the formation of high density, highly ordered planar stripes propagating at a speed of order $w_0$ through a low density,
disordered background. It is important to stress the properties of these stripes: i) they are not a pattern in that the width and spacing of the stripes is not fixed but rather determined by initial conditions and domain size - they are instead solitonic wavefronts;  ii) the  propagation speed $c$ is of the order of $w_0$ and weakly dependent on the base density of the system; iii) The bands are bounded by two sharp fronts - a leading, narrow boundary layer and a wider and more slowly decaying trailing boundary layer (see Fig.~\ref{stripe-structure}). This phenomenology is the same as that found in ~\cite{bertin2009}.

\begin{figure}[tbp]
\includegraphics[width=8cm]{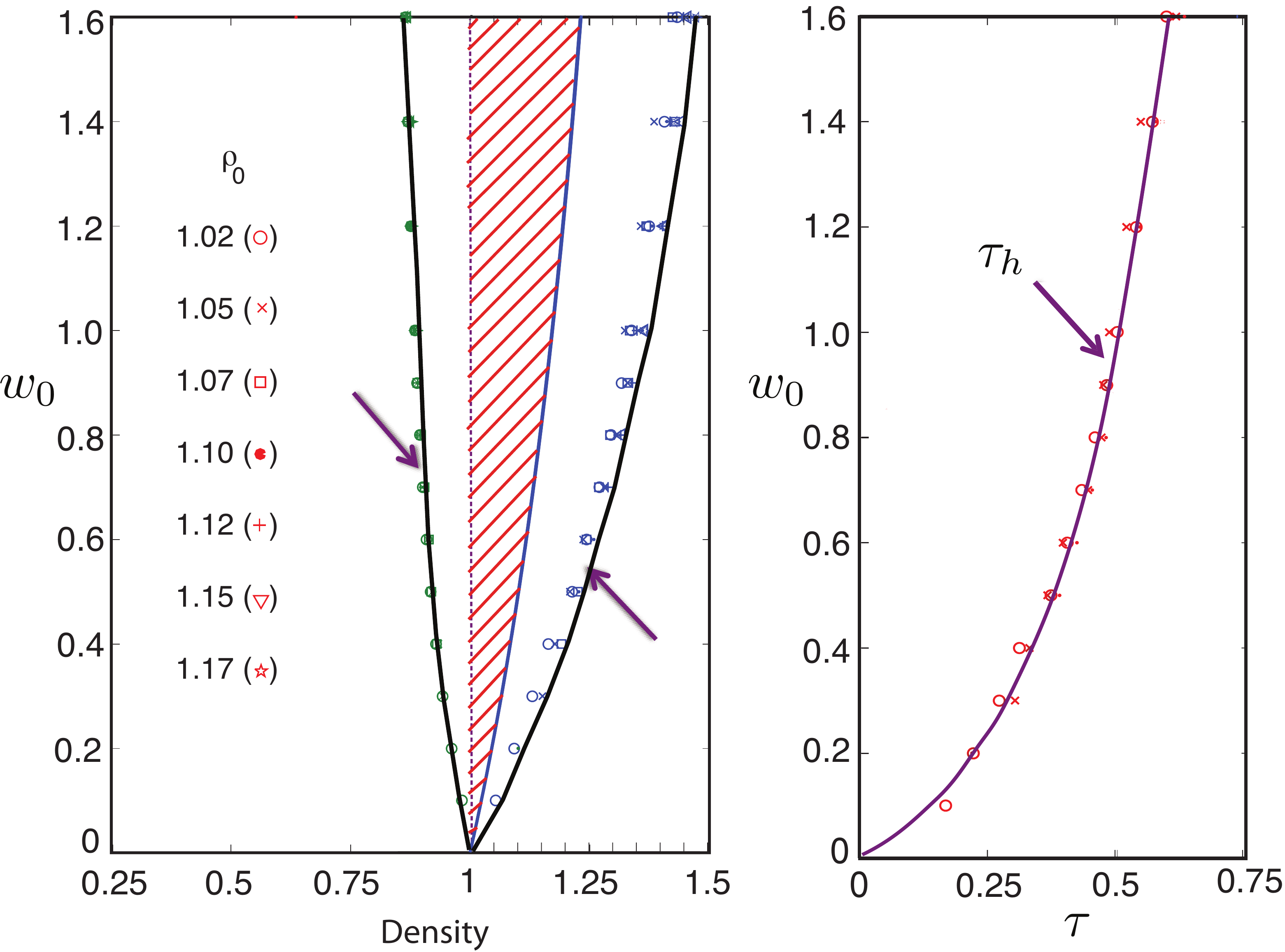}
\caption{Illustration of the phase separation in the stripe state.
\emph{Left Panel}: Computational results obtained at $T=10^{4}$ for a $128 \times 128$ domain
with $\protect\lambda=0$. The shaded region indicates the domain in parameter space when the polar state is unstable due to the convection mediated density instability. The density in the propagating stripe $\protect\rho_{\mathrm{h}}$ and the low density background $\protect\rho_{\ell}$ for various values of the base density and $w_0$ are shown. The lines are a guide to the eye. \emph{Right Panel}: Plot of the polarization $\tau_h$ as a function of $w_0$ from the numerical simulation. The solid line is the analytical prediction $\frac{\rho_{h}(w_0)-1}{\rho_{h}(w_0)+1}$.}
\label{phase-separation}
\end{figure}

\textbf{Mechanism.} The propagating stripe state exists even when we set $\lambda$ to zero. In addition, the onset of the striped state follows closely the neutral stability line associated with the convection mediated density instability determined by Eq.~\ref{L-inst}. Thus, a minimum dynamical model for the emergence of this structure is given by the two coupled equations $\partial_{t}\rho = -w_0 \partial_{x} \tau$ and
$\partial_{t} \tau = -(a_2+a_4{\frac{\tau^2 }{\rho^2}}) \tau -w_0
\partial_{x}\rho $. The two coupled equations are effectively equivalent to a wave equation where the homogeneous nonlinear terms from the polarization equation provide the dispersion that generates the solitary wave structure. After transforming to a co-moving frame, the equations can be reduced to quadrature to determine the speed of propagation and the profile of the wavefront. Such an analysis has already been reported in~\cite{bertin2009}. Here, we present a complementary point of view. As stated earlier, the convection mediated density instability occurs in the vicinity of $\rho_c$. The polar ordered state is re-stabilized at higher densities. In this unstable regime, the dynamics of the system essentially yields a phase separation into a high density ordered state and a low density disordered state, both of which are now stable (see Fig.~\ref{phase-separation}). The degree of order in the stripe is precisely $\tau_h=((\rho_h-1)/(\rho_h+1))^{\frac{1}{2}}$, i.e., the value predicted by the mean field theory for a state of density $\rho_h$. The propagating nature of the ordered state results in the robust concentration waves observed in the numerical solution. Phase separation is also observed in active nematics~\cite{SRShradha} which are also self-regulating in nature, although via mechanism different from polar convection.

\textbf{Experimental Realizations.} Propagating concentrations waves have been  observed in dense actin motility assays~\cite{Schaller} and in self-chemotactic bacterial suspensions~\cite{SilberzanPNAS}. For the actin system, the numerical modeling described in Ref.~\cite{Schaller} yielded the conclusion that local polar aligning interactions among the filaments are necessary for the emergence of the propagating waves. These interactions need not be medium mediated and could arise from the fact that aligned actin filaments have very anisotropic friction constants for sliding along the direction of alignment when compared to sliding against it~\cite{zvonimir}. This conclusion indicates that the propagating waves seen in actin motility assays are indeed the propagating stripes obtained in the present model of polar self-propelled particles due to the dynamic self-regulation of the collective motility. Pattern formation in bacterial suspensions, in contrast, are  controlled by the presence of gradients of signaling chemicals and nutrients and have been traditionally modeled in terms of coupled reaction-diffusion equations. Recent work has shown that some patterns in chemotactic bacteria can be explained by accounting for the fact that the bacterial motility depends on density \cite{CatesPNAS}. In the present context, we want to focus on the  so-called {\em Keller-Segal bands}, extensively studied in the bacteria literature. They arise from self-chemotaxis and are the predominant emergent structure in convection free geometries \cite{SilberzanPNAS}. The alignment interaction among bacteria in these systems scales with the concentration of the chemoattractant, which is in turn proportional to the concentration of bacteria. In this respect  the alignment interaction is similar to the model considered here, where the local ordering tendency is controlled by $a_2$ that depends on the concentration of self-propelled particles. There is, however, an important difference between the two systems in that  in the experiment of Ref.~\cite{SilberzanPNAS} and related studies, the direction of propagation of the bands is set by the nutrient gradient in the microchannel, hence the rotational symmetry is externally broken. In the model considered here, in contrast, the symmetry is broken spontaneously. However, we note that the fact that a symmetry is spontaneously broken is encoded in the dynamics of the associated Goldstone mode and the analysis here shows that the Goldstone mode plays no role in the dynamics that give rise to the concentration waves. Hence, the Keller-Segal bands are also another realization of dynamic self-regulation in self-propelled particle flows.

\subsection{Stationary Asters}

\begin{figure}[tbp]
\includegraphics [width=7cm] {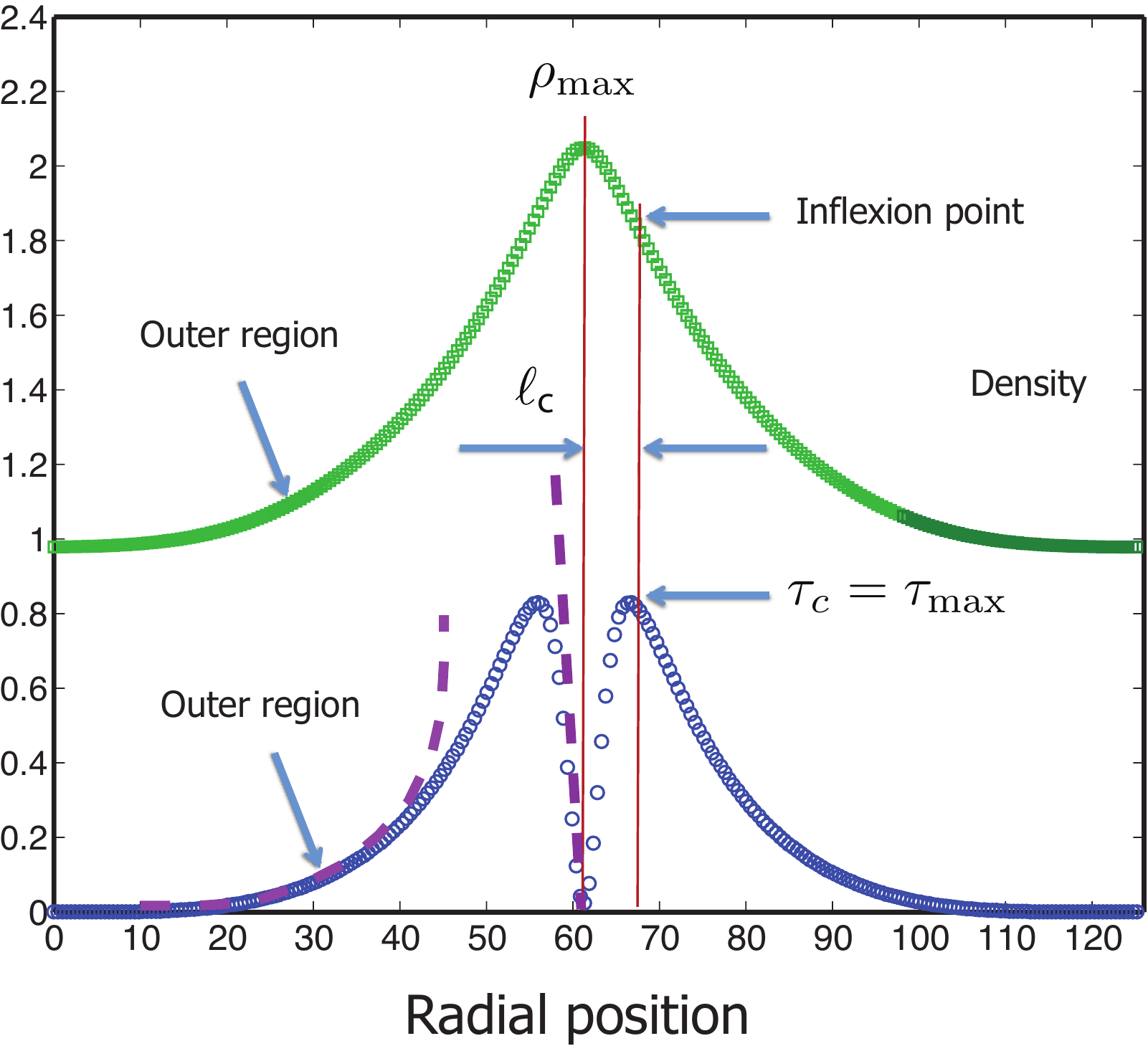}
\caption{Radial density and $|\btau|^2$ profiles of an aster  for $\protect\rho_0 = 1.07$, $\protect\lambda = 0.8$ and $%
w_0=0.05$. The system
size is $128$ x $128$ and the equations have been integrated up
to $10^{4}$ diffusion times - these profiles
remained constant even after integrating to $5$ x $10^{4}$ diffusion times. The
density field $\protect\rho(r)$ exhibits a point of inflection;
at the same radial position the order parameter $|\btau|$ attains a maximum value $\protect%
\tau_{\mathrm{max}}$. The aster is characterized by a core of size $\ell_{%
\mathrm{co}}$ and a region where both density and polarization decay over a characteristic length scale $\ell_\infty$. The dashed curves (purple online) are the theoretical result obtained by the asymptotic analysis based on  Eq.~\eqref{asterAsymp} and described in the text.}
\label{aster1}
\end{figure}

We now focus on the other  inhomogeneous stationary state obtained in the linearly unstable region of parameters, namely a stationary aster. For $\lambda/w_0\gg 1$ the numerical solution of the nonlinear equations yields asters or $-1$ topological defects with radially symmetric profiles of
both density and orientational order. Other authors have obtained stationary asters in models with uniform density~\cite{Kardar,Sankararaman}, where the aster is strictly a defect in the  order parameter. Here in contrast, it is also a region of high concentration. A typical aster is depicted in Fig.~\ref{aster1}. The density
field $\rho$ is a radially symmetric function of $r \equiv |\mathbf{r}|$ about the center of a $-1$ defect located for convenience at the origin of the coordinate system. The
density is a maximum at $r =0$ and decays exponentially far from the core to a value slightly below
the critical density for the onset of ordering, $\rho_c=1$.  Unlike the density field, the order parameter,
$\tau(r)$ is non-monotonic. Starting with a zero value at the core, the
order parameter increases almost linearly to a intermediate maximum value $\tau_{
\mathrm{max}}$ and decays exponentially to zero
far from the aster's core. The point of maximum $\tau$ also corresponds to a point of inflection in the density profile.

We characterize the aster by two length scales, the size $\ell_{\mathrm{co}}$ of the aster core as  defined as the distance from the aster's center to the point where $|\tau|$ reaches its maximum and the density has an inflection point, and  the length scale $\ell_{\infty}$ characterizing the exponential decay of both density and polarization. This second length can be thought of as the size of the aster. Both length scales are only weakly dependent on the interaction strength $\lambda$ for fixed $w_0$. For fixed $\lambda$ we find that both  $\ell_{\mathrm{co}}$ and $\ell_{\infty}$ decrease as a function of $w_0$ while the maximum density and the maximum $\tau$ increase. In other words, for fixed interaction strength larger self-propulsion speed yields denser, tighter asters. The behavior of density and order parameter far form the center of the aster can be obtained from a simple asymptotic analysis of the dynamical equations as the nonlinearities do not play a significant role in determining the asymptotic behavior. Far from the core, assuming a radial dependence of both the density and the order parameter, and using the fact that $\tau \rightarrow 0$ at infinity, we can estimate the magnitude of the polarization by considering the steady state equation to linear order in small deviations from the asymptotic values $(\rho,\tau) = (\rho_{\infty} \leq 1,0)$
\begin{equation}
r^2 {\frac{d^2\tau }{d r^2}} + r {\frac{d\tau }{dr}} -
[r^2(1-\rho_{\infty} + w^2_0) +1] \tau = 0.
\label{asterAsymp}
\end{equation}
The solution to the above equation is a Bessel's function of the second kind, that decays exponentially as ${\mathrm{e}}^{- {\frac{r }{\ell_{\infty}}}}$ with the length scale $\ell_{\infty} \sim
(1-\rho_{\infty} + w^2_0)^{-{\frac{1 }{2}}}$, consistent with the trends identified in the numerical work. The behavior near the  core is determined by a complicated interplay between the nonlinearities and cannot be investigated analytically.

\begin{figure}[tbp]
\includegraphics[width=7cm] {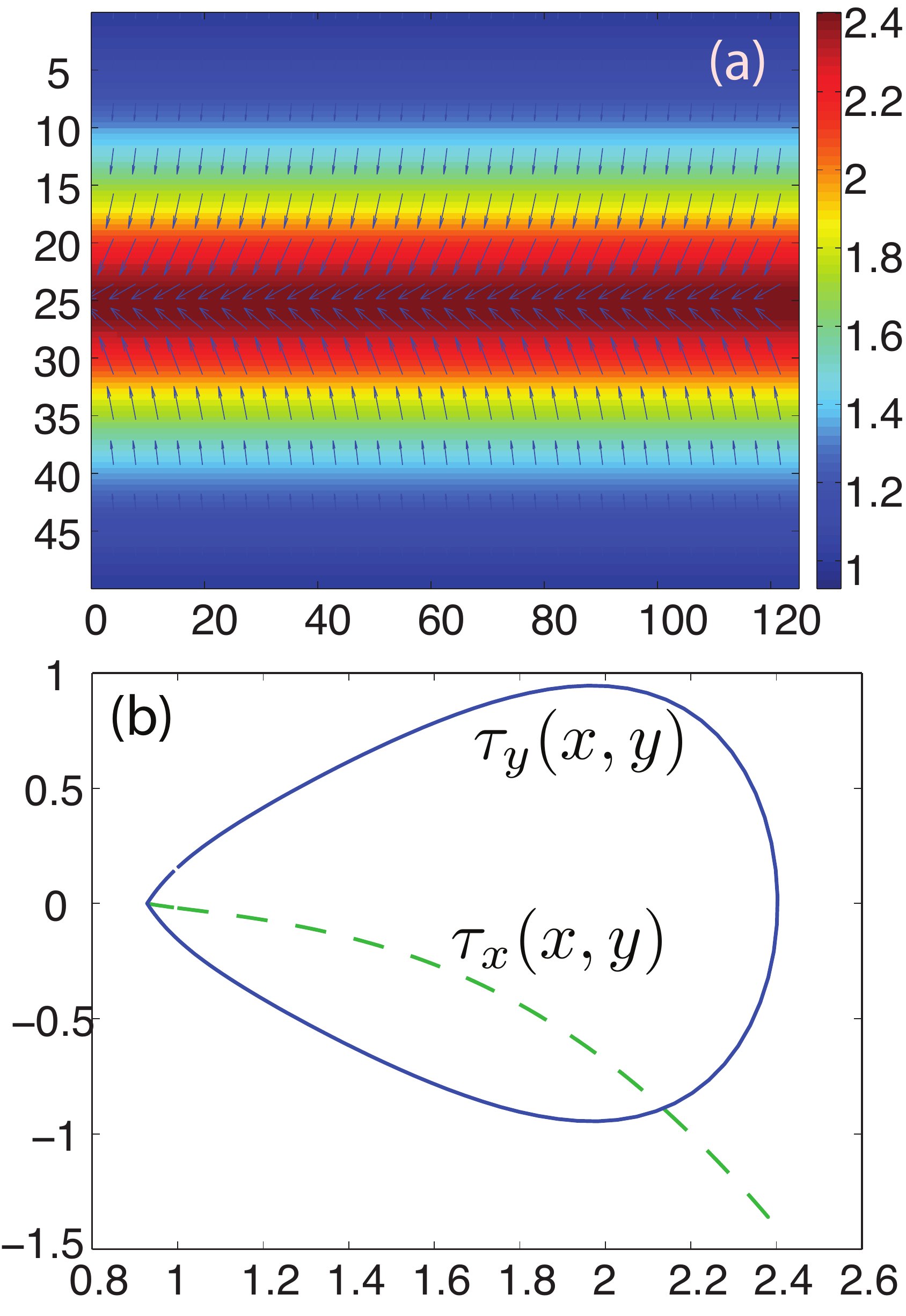}
\caption{Steady inhomogeneous phase-separated structures, we term streamers,  obtained at fixed domain size $128
\times 128$ when $\protect\lambda_1=0$, $\protect\lambda_2=\protect\lambda%
_3=0.8$, $w_0=0.1$ and $\protect\rho_0=1.05$. The
high density (red) region is also highly polar and ordered with the
direction of polarity along the neutral direction. In (b) we quantify this
by plotting the two components of $\btau$ as a function of $y$. Note that $%
\protect\tau_x$ (green dashed line) is a single valued function of $y$ while $\protect\tau_y$ (blue solid line)
changes sign as we traverse the phase separated ordered region from one side
to the other.}
\label{streamer}
\end{figure}

\textbf{Mechanism.} The basic mechanism underlying aster formation is the splay-induced negative compressibility discussed in the context of the linear stability analysis. We stress, however, that the self-regulating nature of the flow is critical for this instability as well in that if we set  $w_0$ to zero, we do not obtain any emergent structures. The system forms asters because this is the only structure that can accommodate both the tendency  to splay and that  to phase separate. Unlike the solitary wave discussed in the previous section, the aster state is not universal and depends on the values of the  parameters  in Eq.~\ref{tau-eq-2}. The self-convection term ($\lambda_1$ in Eq.~(\ref{tau-eq-2})) is critical for the formation of this structure. When we set this term to zero we find that the system develops streamers instead, a phase separated polar ordered state where, in contrast to the solitary wave state, the polarization is orthogonal to the direction of the spatial gradients (see Fig.~\ref{streamer}).

\textbf{Experimental Realizations.}  Aster formation has been seen in purified cell extracts of microtubules and associated motor proteins, such as those studied in Ref.~\cite{nedelec,nedelec2,surrey}. In these controlled in vitro systems, with known concentrations of only a few types of motor proteins, aster formation can be understood theoretically using both continuum models \cite{Kardar,Sankararaman,Aranson} and simulations \cite{surreyTheory}. Understanding the origin of aster-type structures in vivo, such as  the formation and splitting of the mitotic spindle upon cell division, is much more challenging as in this case the process is controlled by a variety of competing nonequilibrium mechanisms, including microtubule polymerization and the on/off dynamics of many different motor types~\cite{spindlelit}.  Our continuum theory where aster formation is controlled by only two competing parameters $\lambda$ and $w_0$ may then provide a useful guidance for the modeling and interpretation of these complex experiments. In particular, one can imagine fitting the spindle structure obtained in experiments on cells where various proteins have been systematically suppressed~\cite{spindlelit,spindlelit2} to our continuum model to identify which proteins directly affect the model's parameters and thereby back out the role played by the protein in the formation of the structure.

\section{Discussion}
In this work, we have considered a generic continuum theory of self-propelled particles in the overdamped limit in two dimensions.  An important new point of our work is the identification of dynamical self-regulation - the fact that the density that controls the amount of order is itself convected by the order parameter - as a crucial mechanism for the formation of emergent structures in these systems. We identify two robust structures in different regions of parameters: traveling solitary waves and stationary asters. We  characterize these structures using numerical solutions of the deterministic nonlinear equations and identify the underlying hydrodynamic mechanisms associated with their emergence.

The primary limitation of this study is that our system is overdamped or ``dry". Momentum is not conserved and there is no coupling between polarization and actual flow. The ``dry" system should be contrasted with an active polar suspension, consisting of active particles in a fluid that mediates hydrodynamic interactions. In the suspension the total momentum is conserved and  important nonequilibrium effects arise from the coupling of the associated flow velocity and local orientational order, as shown in recent work~\cite{Marenduzzo2010,Giomi2010}. A second limitation is the one elastic constant approximation that identifies the energy cost for bend and splay. Theoretical work on models of suspensions of cytoskeletal filaments has shown the importance of retaining different elastic constants for bend and splay to understand the emergence of vortices and spirals \cite{Sekimoto,Joanny}. This physics is missing in the present model.

Finally, we note that aster formation has also been observed in related continuum models when the sign of pressure gradients tends to favor the formation of high density regions~\cite{Zimmermann,Aranson}. In contrast, asters were not found in Refs.~\cite{bertin2009,Mishra2010}, even though the continuum equations considered there have the same structure as those analyzed here. The reason for this is that Refs.~\cite{bertin2009,Mishra2010} study the continuum model obtained by systematic coarse-graining of specific microscopic models  (the Vicsek model and a collection of self-propelled hard rods, respectively) and use the microscopic expressions for the parameters  obtained  in such derivations. In both cases the microscopic calculation yields $\lambda\sim w_0^2$. In other words $\lambda$ is not an independent parameter and cannot be tuned to the large values required to obtain asters.

The simplicity and generic nature of the theory considered in this study has enabled us to highlight the role of dynamic self regulation and to show that the mechanism is universal and does not depend on the microscopic physics, in  contrast to closely related albeit system specific studies such as those in Refs. \cite{Zimmermann, Kruse}.

\subsection*{Acknowledgements}
AG, MFH and AB acknowledge support from the Brandeis-MRSEC through NSF DMR-0820492, and the HPC cluster at Brandeis for computing time. MCM was supported by the NSF on grants DMR-0806511 and DMR-1004789.


\end{document}